\documentclass[12pt,preprint]{aastex}
\newcommand{\hh}{\rm{^2}H}
\newcommand{\dd}{\rm d}
\begin{document}
\title{Light Element Synthesis in High Entropy Relativistic Flows Associated 
With Gamma Ray Bursts}

\author{Jason Pruet}
\affil{Lawrence Livermore National Laboratory, L-414, P.O. Box 808, 
Livermore, CA 94551}
\author{Shannon Guiles and George M. Fuller}
\affil{Department of Physics, University of California, 
San Diego, La Jolla, CA 92093-0319}

\begin{abstract}

We calculate and discuss the light element freeze-out nucleosynthesis
in high entropy winds and fireballs for broad ranges of
entropy-per-baryon, dynamic timescales characterizing relativistic
expansion, and neutron-to-proton ratios. With conditions
characteristic of Gamma Ray Bursts (GRBs) we find that deuterium
production can be prodigious, with final abundance values ${^2}{\rm
H/H}\gtrsim 2\%$, depending on the fireball isospin, late time
dynamics, and the effects of neutron decoupling-induced high energy
non-thermal nuclear reactions. This implies that there potentially
could be detectable local enhancements in the deuterium abundance
associated with GRB events.
\end{abstract}

\keywords{gamma rays:bursts---nucleosynthesis---stars:neutron}
\section{Introduction}

In this paper we show that the high entropy relativistic flows thought
to accompany catastrophic stellar endpoint events like binary neutron
star mergers \citep{schramm} and collapsars \citep{woosley}
may be a source of interesting light element synthesis.  In
particular, we find that under the right circumstances a significant
fraction of the relativistic ejecta can be turned into ${^2}{\rm
H}$. Though this cannot be a cosmologically significant ${^2}{\rm H}$
source on account of the small mass and presumed low event rates
associated with, e.g., GRBs, it could represent an appreciable {\it
local} enhancement. Because the collisionless shocking (and subsequent
synchrotron emission) of ultra-relativistic winds is thought to give
rise to observed GRBs, deuterium production in these winds raises the
possibility of a nucleosynthetic signature of the GRB environment.

The general nucleosynthesis associated with rapid adiabatic expansion
and freeze-out from Nuclear Statistical Equilibrium (NSE) at high 
entropy-per-baryon, $s$, was first considered in the landmark paper
by \cite{wag} (hereafter WFH). Those authors concentrated on the environments 
associated with exploding supermassive objects, where $s/k_b\sim 10^3$,
and on Big Bang Nucleosynthesis (BBN), where $s/k_b\approx10^{10}$. In both 
supermassive objects and BBN the conditions are expected to be proton-rich
(preponderance of protons over neutrons) and the characteristic expansion 
timescale large (e.g., $\tau_{\rm dyn}\sim 100{\rm s}$ for BBN). The physics
of relativistic outflows and potential nucleosynthesis in these sites 
has also been the subject of some recent attention 
\citep{ultra,thomps,otsuki}.

Here we extend the WFH study. We consider freeze-out from NSE over a wide 
range of entropy-per-baryon spanning that in WFH, and a range of 
neutron-to-proton ratios, all the way from proton-rich to neutron-rich.
NSE freezeout in these scenarios is calculated for expansion timescales 
ranging from those appropriate for relativistic flows from compact objects
to those associated with BBN ($10^{-6}{\rm s}<\tau_{\rm dyn}<100{\rm s})$.

We find that the general nuclear physics features of BBN are a 
recurring theme throughout all of these parameter ranges. In particular, 
${^2}{\rm H}$ synthesis can be significant for relativistic flows. The 
expected possibly neutron-rich conditions of a GRB fireball are, in a sense, 
the isospin mirror of those for BBN. However, we also show that it is not 
sufficient to simply consider only the NSE freeze-out nucleosynthesis
in some parameter regimes appropriate for, e.g., GRB fireballs. In fact, 
the dynamics of these fireballs can differ dramatically on a microscopic
scale from the conditions treated by WFH.

In particular, for initially neutron-rich material, the few protons accelerate
with the relativistic photon and $e^{\pm}$ pair fluid to a very high Lorentz
factor ($\gamma\sim10^2-10^3$). In this scenario, there will be high energy 
($\sim {\rm GeV}$) collisions of protons on ``left-behind'' 
neutron stragglers \citep{fpa}. It has been shown that high energy 
collisions of this kind can result in significant destruction and/or 
production of light nuclei \citep{dim}. We show below that these 
non-thermal nuclear reactions can modify significantly the simple 
BBN-like NSE freeze-out abundances of ${^2}{\rm H}$ and other species.
In the right conditions, ${^2}{\rm H}$ number fraction yields can 
approach $Y_{\rm D}\sim 10\%$. This is a staggeringly high yield given 
the known, and small, primordial ${^2}\rm H$ abundance. 

The deuteron is a particularly interesting nucleus because it has a
binding energy of only $\approx$2.2 MeV and is notoriously fragile and
difficult to synthesize. At present it is thought that essentially all
of the deuterium in the universe is primordial. Estimates of the
primordial $\hh$/H come from measurements of the 82 km/s isotope shift in
the Lyman $\alpha$ line in a handful of high redshift Lyman limit
clouds.  These are clouds that intersect our lines of sight to distant
QSO's. The values of $\hh$/H inferred from these clouds are consistent
with each other and average to $\hh/H\sim 3 \cdot 10^{-5}$
\citep{omeara}. Because the deuterium yield in the big bang is a
sensitive function of the entropy per baryon (or alternatively baryon
to photon ratio $\eta$), the observed $\hh$/H places important constraints
on conditions during the epoch of BBN, and gives a determination of
the baryon rest mass contribution to closure (scaled by the Hubble
parameter in units of 100 ${\rm km s ^{-1} Mpc^{-1}}$), $\Omega_b
h^2\approx 0.02$.

Estimates of the present day interstellar medium values of ${^2}{\rm
H/H}$ in different directions from the UV Lyman $\alpha$ transition
are also consistent with each other and average to about ${^2}{\rm
H/H}\approx 1.5\cdot 10^{-5}$
\citep{mcullough}, in accord with the idea that stellar processing
results in a net destruction of deuterium. There have also been a
number of careful radio searches for the hyperfine transition in
deuterium. So far these have only yielded upper bounds about a factor
of four higher than the abundance measured by UV
satellites. Nonethless, radio searches for pockets of ${^2}{\rm H}$ in
supernova remnants may represent a promising avenue for detecting
evidence of relativistic flows. Ultimately whether or not deuterium
from relativistic flows is detectable depends on the interaction of
the flow with the ISM. We do not calculate this here.

\section{ Nucleosynthesis in relativistic flows}

The evolution of a hot plasma is governed by the initial radius $R_0$
of the plasma, the initial temperature of the plasma $T_0$, and the
entropy per baryon $s/k_b$ in units of Boltzmann's constant.
Initially the plasma accelerates under the influence of thermal
pressure. For optically thick flows the acceleration ceases when the
thermal energy of the fireball is converted to kinetic energy. Energy
conservation gives the final Lorentz factor $\gamma_f$ of the fireball
as 
\begin{equation}
\gamma_f=E_{0}/M.
\end{equation}
Here $E_{0}$ is the total initial energy (thermal plus rest mass) in
the fireball and $M$ is the mass of the baryons confined to the
fireball (in this and in what follows we set $\hbar=c=1$). In this
paper we will focus on the case where the initial plasma is baryon
poor, so $\gamma_f > 10$.  Non-relativistic or mildly relativistic
outflows with modest entropy per baryon ($s/k_b \sim 100$) can give
rise to interesting heavy element synthesis
\citep{meyer92,wilsonandmathews,takahashi}, but are relatively uninteresting as
far as A$<$8 nuclei are concerned. In terms of $\gamma_f$ and the
initial temperature, the entropy per baryon in the fireball is

\begin{equation}
  s/k_b\approx \gamma_f {60 \over 45} {m_p \over T} \approx  1250
  \gamma_f({1 {\rm MeV} \over T_0})\approx 10^4-10^6.
\end{equation}

Here $m_p$ is the proton rest mass. At these entropies, and for $T_0
\gtrsim 0.2 {\rm MeV}$, nuclear statistical equilibrium determines
that initially the fireball consists of free neutrons and protons. The
neutron to proton ratio in these relativistic ejecta remains an open
and important question because it may be a probe of conditions near
the fireball source and so help pin down the central engine. When weak
processes are rapid compared to the expansion timescale, n/p will come
to equilibrium with these processes at n/p$\sim 1$ for T$>$ a few
MeV. The rate of electron(positron) capture on free protons(neutrons)
is $\sim 5 (T/{\rm 1 MeV})^5 s^{-1}$. This means that $e^{\pm}$ capture will
influence n/p if $(T/{\rm 1MeV})^5>10^4(10^6{\rm cm}/R_0)$. (Below we
will identify $R_0$ with the dynamic timescale.)  This result implies
that weak processes may or may not influence n/p depending on the
details of the fireball source. For relativistic winds from NS-NS
mergers, for example, neutrino (and $e^{\pm}$) processes are too slow
to influence n/p, so that the composition of the outflow mirrors the
composition of the material ablated/ejected from the neutron stars
\citep{ultra}.  Rough estimates of n/p from different stellar and
compact object sources range from $n/p\approx 10$ for NS-NS mergers to
$n/p\approx 0$ for flows from supermassive object collapse. At any
rate, this is one of the few environments where free neutrons may be
present.  We will treat n/p as an unknown free parameter in this work.

The acceleration of the initially hot plasma has been investigated by
a number of authors. To a good approximation the Lorentz factor and
temperature in the plasma evolve simply as $\gamma=R/R_0=T_0/T$ during
the acceleration stage of the fireball's evolution and as
$\gamma=\gamma_f$ for $R>R_0\gamma_f$ \citep{pac2}.  These scaling
relations imply that an observer comoving with the fireball sees the
temperature evolve as $T=T_0 exp(-tR_0)$. This determines the dynamic
timescale as $\tau_{\rm dyn}=R_0^{-1}$. Compact objects typically have
radii of order $10^6$ {\rm cm} implying $\tau_{\rm dyn} \sim 10^{-5}
s$. The dynamic timescale for flows from jets breaking out of massive
stars during stellar collapse may be several orders of magnitude
larger \citep{woosley}. The jet evolution in these collapsars is too
complicated to allow reliable statements about nucleosynthesis. The
validity of a simple picture for the jet evolution and associated
nucleosynthesis depends on the extent of the interaction between the
outgoing jet and the stellar envelope.

As the relativistic flow expands and cools, NSE begins to favor bound
nuclei. As long as the reactions responsible for keeping nuclei in NSE
are fast compared to the expansion timescale, the nuclear abundances
will track statistical equilibrium abundances. Once the reaction rates
become slow compared to $\tau_{\rm dyn}$, the material freezes out of
NSE and breaks into successively smaller clusters of nuclei in Quasi
Statistical Equilibrium (QSE). Below we will calculate freeze-out
nucleosynthesis yields in relativistic fireballs using the code
developed by Kawano for BBN \citep{kawano}. First, however, we
illustrate some of the salient aspects of nucleosynthesis in these
relativistic fireballs by using a simplified toy reaction network.

Suppose for a moment that deuterium were the only possible bound
nuclear species (i.e. flow to heavier nuclei is prevented). In this
case the evolution of the deuterium number fraction $Y_D$ ($Y_X$ is
defined as the number density of nucleus $X$ divided by the number
density of baryons) is given by

\begin{equation}
{dY_D \over dt}=\rho_b Y_p Y_n \langle \sigma v \rangle_{\rm
pnD\gamma} - Y_{\gamma} Y_D \lambda_{\gamma} \approx \langle \sigma v
\rangle_{\rm pnD\gamma} (\rho_b Y_p Y_n -5\cdot 10^{9}
T_9^{3/2}exp(-Q/T_9) Y_D)
\end{equation}

Here $\langle \sigma v \rangle_{\rm pnD\gamma}\approx 4.7\cdot10^{4}
{\, \rm cm^3/(s mol)}$ is the reaction rate for ${\rm
n(p,\gamma)\dd}$, $T_9$ is the temperature in units of $10^9K$, and
$Q\approx 2.2$ MeV is the binding energy of the deuteron. To a first
approximation, $Y_D$ tracks the equilibrium value, $Y_D^{\rm NSE}$,
until the destruction rate $\lambda_\gamma Y_\gamma \approx 5\cdot
10^9 exp(-Q/T_9)\langle \sigma v \rangle_{\rm pnD\gamma} $ falls below
the expansion rate. (The expression for the reverse rate follows from
the formalism developed by \cite{fcz}.) At this time, $Y_D$ freezes
out at approximately $Y_D^{\rm FO}\approx Y_p Y_n T_9^3
(\tau_{-5}/s_5)$.  Here $\tau_{-5}$ is the dynamic timescale in units
of $10^{-5}s$ and $s_5$ is the entropy per baryon in units of $10^5$.
This result implies that for $(\tau_{-5}/s_5)> 10^{-1}$, i.e. for
essentially all reasonable fireball parameters, there is time to form
interesting abundances of $\hh$. In this case the final $\hh$
abundance is limited by the flow to heavier nuclei.

To estimate the effects of flow to heavier nuclei we employ the code
developed by Wagoner (WFH) and subsequently modified by Kawano. 
This code contains a reaction network for 26 light
nuclei and rates for the relevant weak processes. The parameters of
the ``Universe'' relevant for BBN are: Hubble constant H=$\tau_{\rm
dyn}^{-1}$, a freely chosen initial n/p (in contrast to BBN where n/p
is initially determined by weak reactions and assumed lepton
numbers/spectra), and a final baryon to photon ratio $\eta\approx
3.6/(s/k_b)$. Note that the $\eta$ relevant for these relativistic
winds is some five orders of magnitude larger than the primordial
value ($\eta\sim 6\cdot 10^{-10}$), and that the dynamic timescale for
relativistic winds can be seven orders of magnitude smaller than the
dynamic timescale of $\sim 100 s$ characterizing the epoch of BBN.

\begin{figure}
\plotone{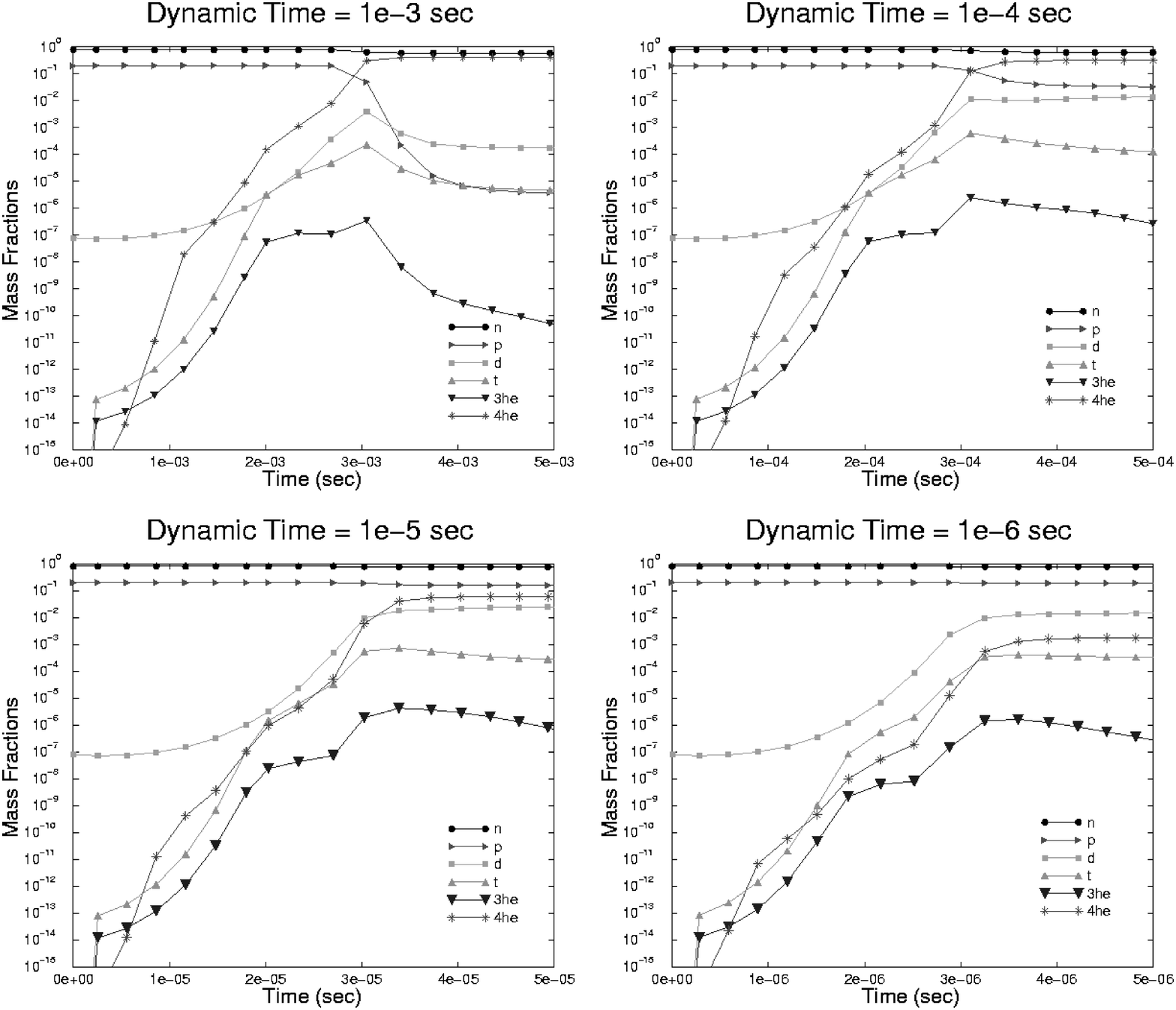}
\caption{Evolution with time of the light element mass fractions for
four different wind cases. For each case the entropy per baryon is $10^5$ 
and $Y_e$ is 0.2. \label{allfig}}
\end{figure}

In Figure \ref{allfig} we show the evolution with time of the mass
fractions of 4 light nuclei for winds with an entropy per baryon
$s/k_b= 10^{5}$ and differing dynamic timescales $\tau_{\rm
dyn}=10^{-3},10^{-4},10^{-5}{\rm and\,\,\,} 10^{-6} s$. This figure
illustrates the effects of the competition between production and
photo-dissociation of $\hh$ for fast outflows, and the flow to heavier
nuclei for slow outflows. For $\tau_{\rm dyn} =10^{-6}s$, the particle
capture reactions out of deuterium are too slow to influence $\hh$/H,
and $Y_{\rm D}$ freezes out at approximately $Y_D^{\rm FO}\approx
1/80$. For $\tau_{\rm dyn}=10^{-5}s$, flow to heavier nuclei,
principally via ${\rm \dd(\dd,n){^3}He}$, ${\rm \dd(\dd,p)t}$, and
${\rm t(\dd,n){^4}He}$ (here t represents the triton, d represents the
deuteron, and the three rates make roughly equal
contributions) becomes important and depletes about 90$\%$ of the
deuterium produced through ${\rm p(n,\gamma)\dd}$.  Of these depleted
deuterons, approximately $0.1\%$ are lost via ${\rm
\dd(p,\gamma){^3}He}$ (or ${\rm \dd(n,\gamma)t}$ for neutron rich
material). The evolution is similar for the slower outflows, with flow
to heavier nuclei becoming increasingly important with time. The
reaction ${\rm \dd(p,\gamma){^3}He}$ begins to dominate as the final
$Y_D$ decreases and the ratio $Y_p Y_D {\rm
\sigma[\dd(p,\gamma){^3He}]}/ Y_D Y_D {\rm \sigma[\dd(\dd,n){^3He}]}
\approx 10^{-5} Y_p / Y_D $ (at $T_9=1$) approaches unity. This is the
case for the $\tau_{\rm dyn}=10^{-3}s$ wind and near this $\tau_{\rm
dyn}$ the final $Y_D$ is very sensitive to small changes in the
entropy and dynamic timescale. This is also the regime most like BBN.

\begin{figure}
\plotone{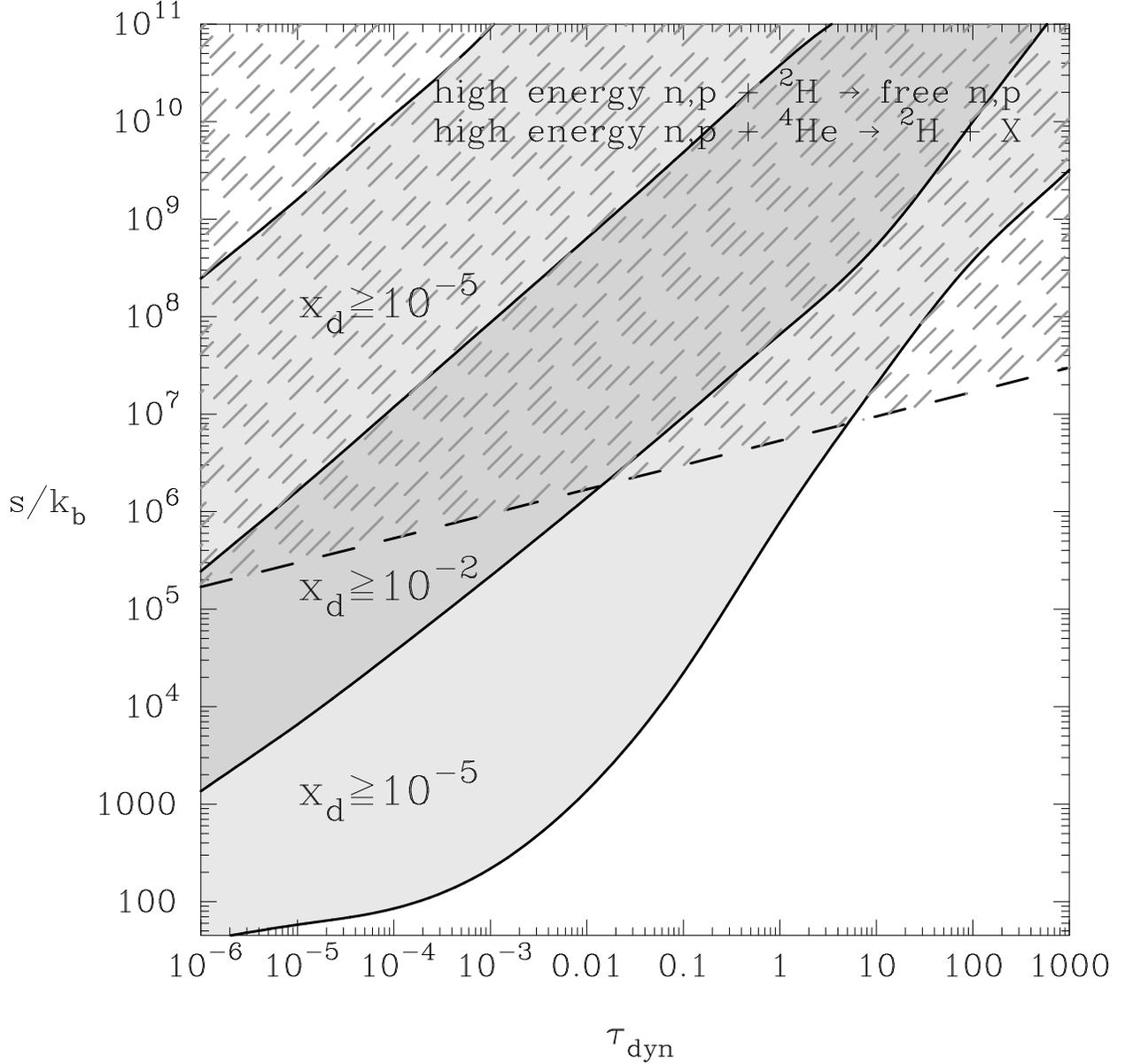}
\caption{ Final deuterium yield as a function of the dynamic timescale
and entropy per baryon characterizing the relativistic flow for
$Y_e=0.5$. The shaded contours were calculated under the assumption
that the neutrons remain well coupled kinematically to the flow. The
upper two lines correspond to $\hh$ production being limited by
photo-dissociation while the lower two lines correspond to $\hh$
production being limited by the flow to $^4{\rm He}$. The thick dashed
line is the neutron decoupling line as described in the text. Above
this line the assumption of kinematically coupled neutrons breaks down
and high energy nucleon-nucleus collisions will change the
nucleosynthesis.\label{fourlines}}
\end{figure}

In Figure \ref{fourlines} we display the final $\hh$ abundance for the
range of entropy and dynamic timescale of relevance to astrophysical
sites. For this figure we assume an initial electron fraction
$Y_e=0.5$.  Note that for expansion timescales comparable to the free
neutron lifetime the final yields are sensitive to the temperature at
which the neutron to proton ratio is specified. We show these very
long expansion timescales only for illustrative purposes. The two
upper contours in Figure \ref{fourlines} correspond to deuterium
production being limited by ${\rm \dd(\gamma,n)p}$, and the two lower
contours correspond to deuterium production being limited by flow out
of $\hh$ to heavier nuclei principally by the strong interactions.

 The thick dashed line in the upper half of the graph delimits the
region where dynamic neutron decoupling occurs.  Dynamic neutron
decoupling occurs because photons and electron/positron pairs dominate
the dynamics of the accelerating flow. Protons and charged nuclei are
coupled to the relativistic light particles electromagnetically, but
free neutrons are uncharged and have a small magnetic moment, so they
are dragged along only by n-p collisions. When these collisions become
slower than the dynamic expansion timescale, the neutrons are left to
coast while the protons accelerate without them. In the absence of
bound nuclei, the condition that dynamic neutron decoupling occurs is
that $\tau_{\rm dyn} n_p \langle \sigma v \rangle _{\rm np } <1$
before the end of the acceleration phase of the fireball's evolution
(here $n_p$ is the number density of protons).  Putting in the scaling
relations describing the evolution of the fireball gives this
condition as $s/10^5>3(\tau_{\rm dyn} Y_p / 10^{-5})^{1/4}$ (e.g.,
\cite{fpa}).

The precise neutron decoupling condition depends on the nuclear
composition of the flow. If, for example, all of the nuclei are locked
into $\alpha$ particles, then neutron decoupling cannot
occur. Roughly, the fraction of nucleons bound in nuclei is small
($\lesssim 10\%$) for $\tau_{\rm dyn} <10^{-4}s$ and entropies along
the neutron decoupling line, and the fraction of nucleons bound into
$\alpha$ particles approaches saturation for $\tau_{\rm dyn}>10^{-4}$
and entropies along the neutron decoupling line.  This implies that
for longer dynamic timescales neutron decoupling will only occur if
the flow is neutron rich, or if the entropy is a factor of a few
higher than that given by the neutron decoupling line in the
figure. 

If the flow is neutron rich, neutron decoupling will result in
high energy (several hundred MeV) neutron-nucleus collisions which
will destroy and synthesize nuclei. These non-thermal reactions are
not described by the Kawano reaction network. The net result of these
collisions depends on the nuclear composition at the time of
decoupling. 

We will consider two instructive limiting cases: (i)
deuteron rich and alpha poor, and (ii) deuteron poor and alpha
rich. Case (i) occurs for short dynamic timescales and for entropies
well above the neutron decoupling line, while case (ii) occurs for
longer dynamic timescales and entropies near the neutron decoupling
line. The final nucleosynthesis is also sensitive to the (unknown)
details of neutron decoupling. In particular, we will see that the
final deuterium abundance is sensitive to the exact number of high
energy interactions suffered by the average nucleus. This number is
certainly between one and a few, but the exact number can be obtained only
with a detailed transport calculation.

Any deuterons present when neutron decoupling occurs will be
destroyed.  Likewise, any $\alpha$ particles present will
be broken apart. Because the branching ratio for $\alpha+({\rm \sim
GeV\,\,\,nucleon}) \rightarrow {^2}{\rm H}+X$ is high, about $50\%$
\citep{dim}, the production of $\hh$ through the spallation of $\alpha$'s 
may
(over)compensate the loss of deuterium in direct collisions. Certainly
for case (i), the $\alpha$ poor case, neutron decoupling will result
in a net decrease in deuterium. For case (ii), where $\alpha$
particles dominate, neutron decoupling will drive $Y_{\rm D}$ to
approximately half of the initial (i.e. at the beginning of neutron
decoupling) value of $Y_{{^4}\rm He}$, provided that the average
nucleus only undergoes one high energy collision. If the average
nucleus undergoes a few high energy collisions, the nucleosynthesis is
somewhere inbetween the one collision case and the fixed point case
discussed by \cite{dim}. 

For example, if initially the material is composed entirely of
$\alpha$ particles and free neutrons, and each $\alpha$ particle
suffers one high energy collision with a neutron, then $\sim 10\%$ of
the material will be converted to $\hh$. Regions in Figure
\ref{fourlines} above the decoupling line where the freeze-out mass
fraction of $\alpha$ particles is large could approximate this
case. If additionally this initial spallation takes place right at the
neutron decoupling point, then none of the deuterium produced from the
fragmenting of $\alpha$ will be destroyed. Only if these conditions
are met can the final $\hh$ yield be as high as $\sim10\%$. 

 Note that
here there is no source of significant post-nucleosynthesis
photo-dissociation of deuterium. This is because on average there will
only be $\sim 100 {\rm MeV}$ per baryon in the form of electromagnetic
radiation coming from the decay of pions produced in inelastic
nucleon-nucleon collisions. This is three orders of magnitude smaller
than the $\sim 100{\rm GeV}$ per baryon coming from the decay of
massive particles discussed by Dimopoulos et al.. In addition, here the
energy injection by $\pi$ and $\mu$ decay is well approximated as
occurring at the decoupling temperature, $T_{\rm D}\approx 0.005(s_5/
\tau_{-5}Y_e)^{1/3} {\rm MeV}$ . This means that, unlike the case
where a massive particle decays with a lifetime of many system dynamic
timescales, the maximum energy of the nonthermal breakout photons does
not keep rising with time.  Interestingly, the recognition that the
branching ratio for $\hh$ production in high energy nucleon-$\alpha$
collisions is so large suggests that relativistic flows might lead to
direct deuterium production via collisions with dense winds or ejecta
surrounding the site of the relativistic wind \citep{fgp}.

\begin{figure}
\plotone{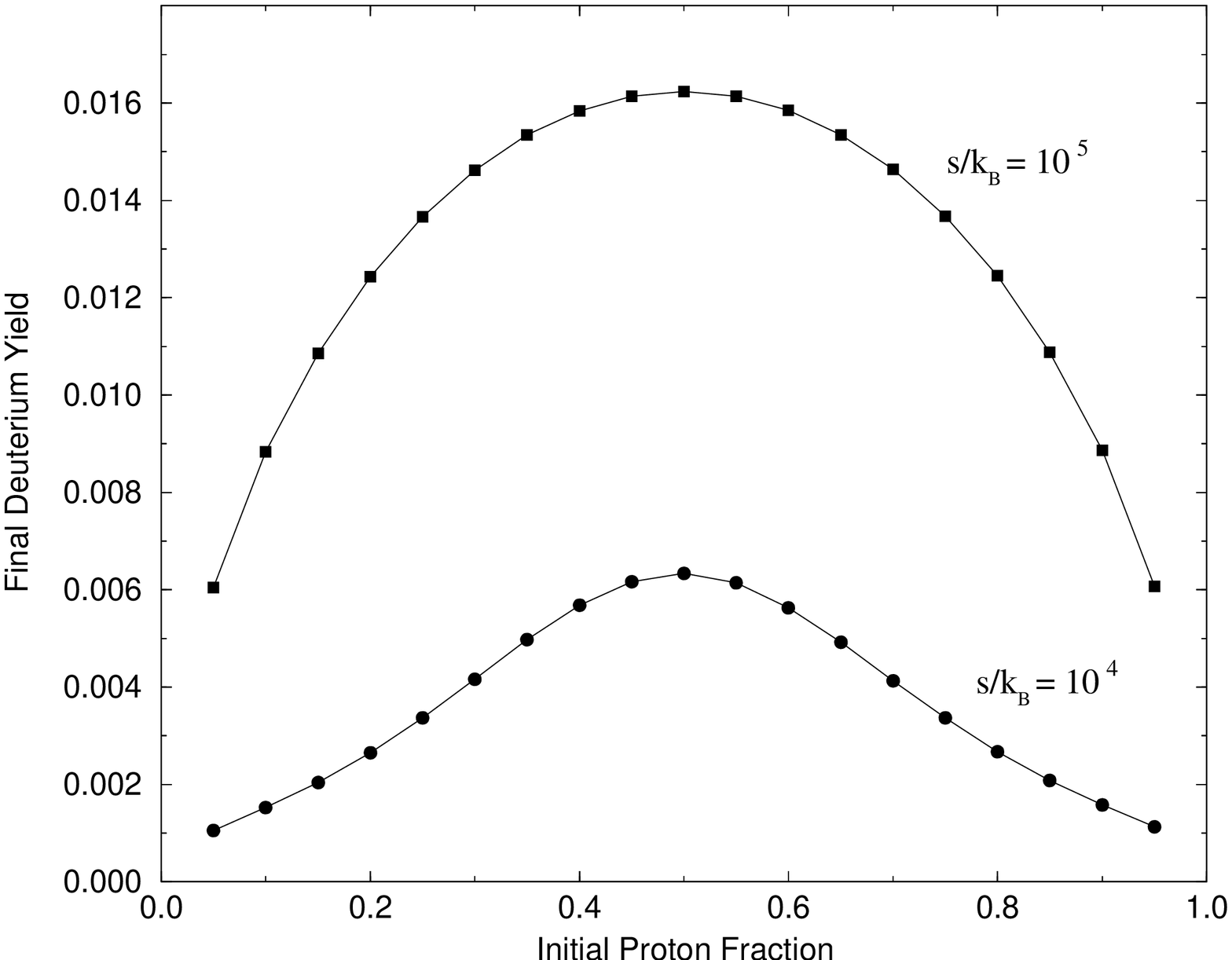}
\caption{ 
Final deuterium yield as a function of the electron fraction. The upper curve
is for $\tau_{\rm dyn}=10^{-5}s$ and $s/k_b=10^5$, while the lower curve is for
 $\tau_{\rm dyn}=10^{-5}s$ and $s/k_b=10^4$.\label{ydvsye}}
\end{figure}

In Figure \ref{ydvsye} we illustrate the effect that changing the neutron to
proton ratio has on the nucleosynthesis. Two illustrative examples are
shown. The upper curve corresponds to a high entropy, fast expansion  and
there is little synthesis of nuclei heavier than ${^2}{\rm H}$. In
this case the final deuterium abundance follows a clean, simple
freezeout from NSE, i.e., the Saha equation: $Y_{\rm D}\propto
Y_pY_n\approx Y_e(1-Y_e)$, where $Y_e$ is the number of electrons per
baryon. The lower line in Figure \ref{ydvsye} corresponds to a lower 
entropy and may be thought of as a convolution of the upper curve
with the reactions ${\rm \dd(\dd,n){^3}He}$ and ${\rm
\dd(\dd,p)t}$. The effect of destruction is less pronounced on the
wings of the curve simply because the deuterium abundance is smaller
and the reactions ${\rm \dd(\dd,n){^3}He}$ (and other comparable
reactions) are slower.

At this point we can address in broad stroke the issue of Deuterium
production in proposed GRB models. For long bursts, observed redshifts
imply a total energy output in gamma rays in GRBs of order
$10^{53}-10^{54}(\Omega/4\pi){\rm erg}$. Here $\Omega$ is the opening
angle subtended by the relativistic outflow. There is some evidence
that long GRBs are typically beamed by a factor of $\Omega/4\pi\approx
1/500$ and that the energy output in gamma rays in long GRBs is
burst-independent and about $10^{51}{\rm erg}$. For short GRBs no
redshifts have been measured and little is known about the energy
scale.  Not all of the energy $E_0$ can be converted to gamma rays and
$E_0$ is larger than the energy in gamma rays by at least a factor of
5-10. An estimate of the energy scale gives the initial temperature in
GRBs

\begin{equation}
T_0\approx 10{\rm MeV}\left[ {E_0 \over 10^{52} {\rm erg}} \left({
4\pi \over \Omega}/ {1 \over 500} \right) {10s \over T_{\rm burst}}
{10^{6} {\rm cm} \over R_0} \right]^{1/4}
\end{equation}

The Lorentz factors invoked to explain GRBs lie in the range $\sim
10^2-10^3$, with the smaller values corresponding to internal shock
models for long bursts and the larger values corresponding to external
shock models for short bursts. This implies entropies per baryon in
the range $10^4- 10^6$. From Figure \ref{fourlines} we see that GRBs,
and relativistic flows in general, associated with compact objects
will be accompanied by substantial deuterium production.

\section{Conclusion and Discussion}

There are three length scales discussed in connection with GRBs. The
first is the size of the central engine ($R_0 \sim 10^6{\rm cm}$ for
compact objects, $R_0\sim 10^{11}{\rm cm}$ for collapsars). This
region has been the focus of many detailed attempts at understanding
the dynamics of the collapsing object giving rise to the relativistic
flow. The second length scale is the scale associated with the
shocking that generates the observed signal, $\gamma_f ^2 R_0$ for
internal shocks or $\sim 10^{15}{\rm cm}$ for external shocks. For
obvious reasons, this scale also has been the subject of many
studies. In this paper we have considered the third, intermediate
scale ($\sim \gamma R_0$) associated with the acceleration of the
thermal fireball. This region is rich in nuclear and particle
phenomena and is the only region sensitive to, e.g., the neutron to
proton ratio in the flow.

We have explored nucleosynthesis for the broad range of dynamic
timescales, neutron to proton ratios, and entropy per baryon which may
be found in these relativistic flows surrounding stellar endpoint
events. For relativistic flows for which the neutrons remain
kinematically well coupled, the nucleosynthesis is similar to the
freeze-out nucleosynthesis discussed in connection with BBN. The final
mass fraction of deuterium can be greater than a few percent, some
three orders of magnitude larger than the primordial value. For flows
for which the assumption of kinematically well coupled neutrons breaks
down, light element synthesis is dominated by high energy non-thermal
nucleon-nucleus collisions. In these latter cases the deuterium number
fraction can be high, $Y_D\lesssim 10\%$. This local overabundance of
deuterium will reside in the shell propagating into the ISM at late
times. Recently, these shells have gained some attention because a
radio identification of a young ($\sim$ 100 yr) non-spherical HI shell
might tell us something about GRBs and their association with
supernovae \citep{pacz}.

Whether or not the deuterium produced in stellar endpoint events is
detectable depends ultimately on the degree of mixing with the
ISM. The total amount of mass in deuterium produced in these events is
${\rm M_D} \approx 10^{-6} {\rm M}_{\odot} (E/10^{52}
erg)(100/\gamma_f)(Y_d/10^{-2})$, where $E$ is the total energy in the
relativistic flow. For HI column densities of order $10^{19}/{\rm
cm^2}$, this deuterium will lead to a factor of two increase in
$\hh$/H over the primordial $\hh$/H as long as it mixes in a volume
less than about a cubic light year. Structures of this size associated
with nearby supernova remnants are seen and resolved with the
VLA. However, even with the most optimistic assumption that every
supernova is accompanied by a relativistic wind, the mechanism
discussed here can only contribute to the {\it overall} present day
deuterium abundance at the level of about 0.1 percent.

It is perhaps frustrating then that such a prodigious yield of deuterium 
has little leverage on the scale of the cosmological abundance. However,
the potentially large $\hh$ production in relativistic flows could possibly
be detectable and, therefore, provide a new probe of, e.g., GRB central 
engine physics and expansion dynamics.

\acknowledgments
 
We acknowledge discussions with K. Abazajian, D. Kirkman, J.  O'Meara,
and S. Woosley.  This work was supported in part by NSF Grant
PHY-0099499 at UCSD and DOE Sci-Dac supernova grants at LLNL and UCSD.

\end{document}